\documentclass[%
 twocolumn,
 amsmath,amssymb,
 aps,
]{revtex4-1}

\usepackage{graphicx}% Include figure files
\usepackage{dcolumn}% Align table columns on decimal point
\usepackage{bm}% bold math
\usepackage{hyperref}% add hypertext capabilities
\usepackage{color}

\begin{document}
\title{Nonlinear Bound States in the Continuum in \\ One-Dimensional Photonic Crystal Slab}
\author{S.D. Krasikov}
 \email{s.krasikov@metalab.ifmo.ru}
 \affiliation{ITMO University, St. Petersburg 197101, Russia}
\author {A.A. Bogdanov}
 \affiliation{ITMO University, St. Petersburg 197101, Russia}
\author{I.V. Iorsh}
 \affiliation{ITMO University, St. Petersburg 197101, Russia}
 
 \begin{abstract}
    Optical bound state in the continuum (BIC) is characterized by infinitely high quality factor resulting in drastic enhancement of light-matter interaction phenomena. We study the optical response of a one-dimensional photonic crystal slab with Kerr focusing nonlinearity in the vicinity of BIC analytically and numerically. We predict a strong nonlinear response including multistable behaviour, self-tuning of BIC to the frequency of incident wave, and breaking of symmetry protected BIC. We show that all of these phenomena can be observed in silicon photonic structure at the pump power of several $\mu$W/cm$^2$. We also analyze the modulation instability of the obtained solutions and the effect of the finite size of the structure on the stability. Our findings have strong implications for nonlinear photonics and integrated optical circuits.      
 \end{abstract}
\maketitle

\section{Introduction}
Bound states in the continuum~\cite{Hsu2016} (BICs) are a special class of localized solutions of wave equations, which have the energy lying in the continuum of the delocalized states. These states may be interpreted as resonant states with infinite quality factor, which originate due to destructive interference of several leaky modes of the system. BICs  are a general feature of wave dynamics, and so may arise for quantum mechanical particles ~\cite{Neumann,Stillinger1975a,Capasso1992}, sound waves~\cite{Parker1967,Koch1983,Evans1994}, water waves~\cite{Ursell1951,Retzler2001,Cobelli2011} and photonic structures~\cite{Bonnet-Bendhia1994,Tikhodeev2002,Lee2012,Sadrieva2017}. The systems supporting optical BICs are usually realized as a two- or one-dimensional periodic photonic structures, such as photonic crystal slabs~\cite{Hsu2013} or patterned photonic wires~\cite{Bulgakov2017}. The \textit{continuum} in this case represents the states which have the tangential component of the wavevector smaller than the total wavevector of the plane wave in surrounding medium at the same frequency. 
The bound states in the continuum in photonic crystal slabs and plasmonic lattices being high quality factor resonant modes already found applications for sensing~\cite{Yanik2011}, filtering~\cite{Foley2014} and lasing~\cite{Hirose2014,Kodigala2017}.

One of the most straightforward possible applications of high quality resonance modes such as BIC is the enhancement of the optical nonlinear effects. The nonlinear dynamics of bound states in the continuum has been explored quite recently in Fabry-Perot cavities~\cite{Bulgakov2010} comprising a nonlinear impurities~\cite{Bulgakov2009}. Moreover, nonlinear bound states in the continuum which are conventionally referred to as embedded solitons have been actively studied for more than decade in nonlinear fibers~\cite{Yang2001}.

The simplest manifestation of the nonlinear dynamics in the resonant photonic structures is the multistable optical response. If a nonlinear system is pumped by an input beam which frequency is detuned from the linear resonant frequency, then the nonlineary frequency shift of the appropriate sign may tune the system in resonance with the pump beam which is the origin of the multistability~\cite{Agrawal1979}. The optical multistability (or bistability as a most simple case) has direct applications for the realization of all-optical switchers.
In real structures, bistability onset is determined by the threshold pump power density which depends on the nonlinear susceptibility of the material and the resonator quality factor.
Ultra high quality factors of the BIC thus suggests low threshold powers for the bistability onset.
\begin{figure}[htbp]
\centering
\includegraphics[width=\linewidth]{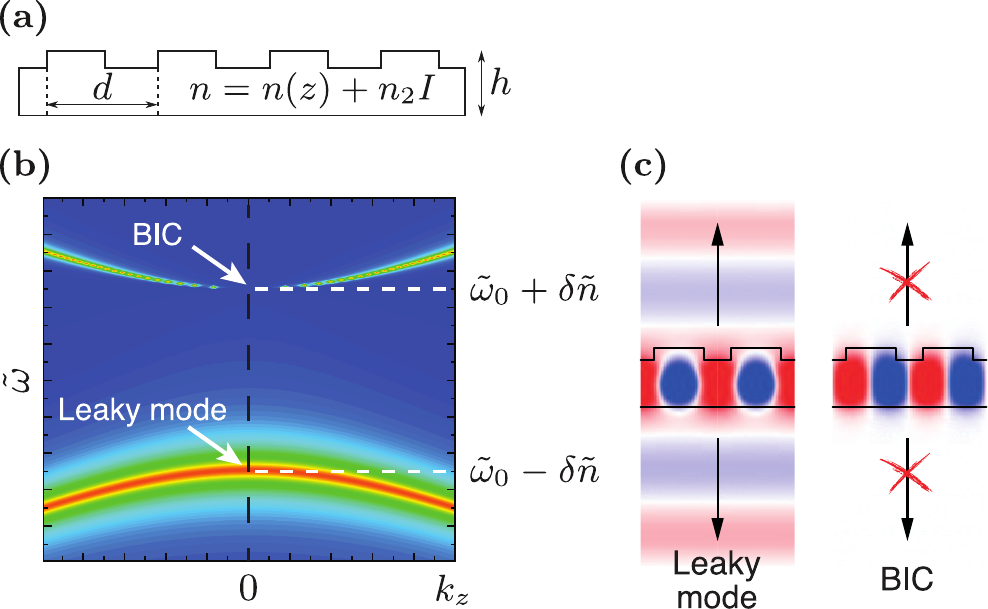}
\caption{(a) Schematic image of the considered system: 1D photonic crystal slab. (b) Map of the reflection coefficient versus in-plane wavevector and frequency. (c) Electric field distribution of leaky (symmetric) mode and bound state in the continuum (anti-symmetric mode).}
\label{fig:system}
\end{figure}

In this work, we study the nonlinear response of a one-dimensional photonic crystal slab with Kerr-type focusing nonlinearity supporting BIC. We analyze the role of BIC in formation of bi-, tri-stable states, and study the modulation instability accounting for a finite size of the photonics structure. We show that the BIC supporting systems allow to achieve strong nonlinear response without cavity at low pump power. 

\section{coupled-mode equations}

The structure under consideration is shown in Fig.~\ref{fig:system}(a). For the sake of simplicity, we focus only on the TE-polarized waves since the problem is reduced to the scalar form in this case.
%It is one of the simplest realization of BIC supporting systems~\cite{Fox1975}. 
%Being ultra high resonant state, BIC manifests itself in the reflectance spectra as a collapse the Fano-shaped resonance at the normal incidence [Fig.~\ref{fig:system}(b)].
We will describe the structure in terms of effective refractive index considering periodicity and nonlinearity as petrubations:
%As seen from Fig.~\ref{fig:system}(a) the structure is a one dimensional grating, which can be treated as a waveguide with periodically modulated  refractive index $n(z)$:
\begin{equation}
	n(z) = n_0 + \delta n \frac{2}{\pi} \sum\limits_j \frac{1}{j} \cos\left(jGz \right) + n_2 I.
	\label{eq:RI_distribution}
\end{equation}
Here $n_0$ is the average refractive index of core layer in the absence of grating, $\delta n$ is the modulation amplitude, $G = 2\pi/d$ is the reciprocal vector of the structure, $d$ is the period of lattice and $n_2 I$ is the term corresponding to the Kerr-type focusing nonlinearity. Both $n_0$ and $\delta n$ are found from the solution of the conventional slab waveguide problem. The geometry of the waveguide is chosen such that, the reciprocal vector $G$ is close to the wavevector of the effective waveguide mode. Thus, the normally incident radiation resonantly excites forward and backward propagating waveguide modes, which are coupled via backscattering. Then we exploit the coupled-mode theory in the simplest form of two-mode approximation~\cite{Kogelnik_1975_Theory_of_Waveguides, Sipe_1994_Propagating_through_gratings,Kogelnik_1971_CME_for_grating}. The electric field can be written as
\begin{align}
	E(z,t) = \left[E_{+}(z,t) e^{i\beta z} + E_{-}(z,t) e^{-i \beta z} \right] e^{-i\omega_0 t} + c.c.
	\label{eq:mode_expansion}
\end{align}
Here $E_{\pm}$ are the amplitudes of the forward and backward propagating modes with the wavevector $\beta$. Implementation of the slow-varying approximation allows to neglect the second derivatives in the wave equation. The terms proportional to $\delta n^2$, $(n_2 I)^2$ and $\delta n\times n_2 I$ are omitted. Moreover, we assume $\omega_0 = ck_0= \beta c/n_0$ and use $\beta = G=2\pi/d$ omitting all terms which have the spatial dependence other than $e^{\pm i \beta z}$.  
With these assumptions, the wave equation can be transformed to a pair of the dimensionless coupled-mode equations:
\begin{align}
	\frac{\partial\tilde{E}_{\pm}}{\partial\tilde{t}} =\mp\tilde{E}_{\pm,\tilde{z}} +i\delta\tilde{n}\tilde{E}_{\mp}+i\tilde{E}_{\pm} \left(|\tilde{E}_{\pm}|^2 + 2 |\tilde{E}_{\mp}|^2 \right).  
	\label{eq:CME}
\end{align}
Here we use the following normalized parameters:
\begin{eqnarray*}
\tilde z&=&\beta z;  \ \ \  \delta \tilde n =\delta n/(2 \pi n_0);  \\ \tilde t &=&\omega t; \ \ \  \tilde E_\pm=E_\pm(n_2/n_0)^{1/2}.
\end{eqnarray*}
Further, we will consider only spatially uniform solutions, and thus drop the spatial derivatives. These equations however should be supplemented though with the pump and decay terms. For that we introduce the vector of the resonant mode components $\mathbf{E}=(\tilde{E}_{+},\tilde{E}_{-})^T$. Moreover, we introduce the vectors of incoming and outgoing fields $\mathbf{S}_{+}$ and $\mathbf{S}_-$. We assume that the incoming radiation is normal to the interface impinging either from the top or from the bottom interface. The energy can leave the system also only normally to the interface from top or bottom. Thus, both 
$\mathbf{S}_{+}$ and $\mathbf{S}_-$ are vectors with size~2. The resulting set of equations can be written as
\begin{align}
&\frac{\partial {\tilde E}_{p}}{\partial \tilde{t}}=i\delta \tilde n\hat{\sigma}_{x,pq}{\tilde E}_j+iT_{pqkl}\tilde E_q\tilde E_k^*\tilde E_l-\hat{\Gamma}_{pq}\tilde E_q+\hat{D}^{T}_{pq}S_{+,q},\\
&\frac{\partial S_{-,p}}{\partial \tilde{t}}=\hat{C}_{pq}S_{+,q}+\hat{D}_{pq}\tilde E_q.
\end{align}
Here, $p,q,k,l=\{+,-\}$, $\hat{\sigma}_x$ is the Pauli matrix, $T_{pqkl}=\delta_{pq}\delta_{kl}(2-\delta_{pk})$ is the tensor governing the structure of the nonlinear response, $\hat{C}=r_{slab}\hat{I}+t_{slab}\hat{\sigma}_x$ is the matrix representing the non-resonant reflection and transmission from the slab ($r_{slab}$, $t_{slab}$ are Fresnel coefficients for the effective uniform slab) and $\hat{\Gamma}á \hat{D}$ are unknown matrices responsible for the decay of the resonant modes and their coupling with incoming and outgoing waves, respectively. In Ref.~\cite{Suh_2004_Temporal_CME} certain relations between $\hat{\Gamma},\hat{D},\hat{C}$ were derived from the energy conservation and time reversal invariance condition. Namely, it was shown that %$ \hat{D}^{\dagger}\hat{D}=2\hat{\Gamma}$ and $\hat{C}\hat{D}^*=-\hat{D}$.
\begin{align}
&\hat{D}^{\dagger}\hat{D}=2\hat{\Gamma},\label{eq:Fancond1}\\& \hat{C}\hat{D}^*=-\hat{D}.\label{eq:Fancond2}
\end{align}
These relations are met even for nonlinear system, since nonlinearity is conservative and does not break the time-reversal symmetry.
%While, in  Ref.~\onlinecite{Suh_2004_Temporal_CME} the Authors considered the linear case only, it can be shown the relations also hold for our nonlinear case, since nonlinearity is conservative and does not break the time-reversal symmetry.
Because of the symmetry of the problem we seek $\hat{D}$  and $\hat{\Gamma}$ in the form: 
\begin{align}
\hat{D}=e^{i\phi_0}\begin{pmatrix}\sqrt{\tilde\gamma} & \sqrt{\tilde\gamma} \\ \sqrt{\tilde\gamma} & \sqrt{\tilde\gamma}\end{pmatrix}; \ \ \ \ \ \ \ 
\hat{\Gamma}=\begin{pmatrix}{\tilde\gamma} & {\tilde\gamma} \\ {\tilde\gamma} & {\tilde\gamma}\end{pmatrix}.
\end{align}
The constant $\tilde\gamma=\gamma/(2\beta^2)$ can be estimated from the consideration that the coupling efficiency is proportional to the refractive index contrast, namely $\gamma\approx k_0^2\delta\tilde{n}^2$. The phase $\phi_0$ is found from the condition~\eqref{eq:Fancond2} and yields $\phi_0 = -1/2i \log(-r_{slab} - t_{slab})$. For the case of a thin waveguide it can be approximated by $\phi_0\approx (n_0k_0 h+\pi)/2$, where $h$ is the thickness of the waveguide. %Matrix $\hat{\Gamma}$ then reads
%\begin{align}
%\hat{\Gamma}=\begin{pmatrix}{\gamma} & {\gamma} \\ {\gamma} & {\gamma}\end{pmatrix}.
%\end{align}
%The constant $\gamma$ can be estimated from the consideration that the coupling efficiency is proportional to the refractive index contrast, namely $\gamma\approx k_0^2\delta\tilde{n}^2$. 
The set of equations is then rewritten as
\begin{align}
&\frac{\partial \tilde E_{\pm}}{\partial \tilde t}=i\delta \tilde  n \tilde  E_{\mp}-\tilde\gamma(\tilde  E_{+}+\tilde  E_{-})+\nonumber \\&i\tilde  E_{\pm}(|\tilde E_{\pm}|^2+2|\tilde E_{\mp}|^2)+\sqrt{\tilde \gamma \tilde I_p}e^{i\phi_0-i\delta\tilde\omega \tilde  t}.
\label{eq:CMT-1}
\end{align}
Here $\tilde I_p = n_2 E_p^2/2$ and $\delta \tilde \omega=\delta\omega/\omega$ is the detuning between the pump beam and $\tilde\omega_0=\omega_0/\omega$.

\section{Self-tuning of BIC and symmetry breaking}
In the absence of pumping and nonlinearity,  system of Equations \eqref{eq:CMT-1} has symmetric and antisymmetric  solutions:
\begin{eqnarray}
\tilde\omega&=&\delta\tilde n; \ \ \ \ \ \  \ \ \ \ \ \ \ \ \ \tilde E_+=-\tilde E_- \\
\tilde\omega&=&-\delta\tilde n - 2i\tilde\gamma; \ \ \ \ \ \tilde E_+=\tilde E_-.
\end{eqnarray}

The antisymmetric solution has no radiation losses and corresponds to the symmetry protected BIC whereas the symmetric solution corresponds to leaky modes [see Fig.~\ref{fig:system}(b),(c)]. In the nonlinear case, amplitude of BIC depends on the frequency:
 %We first analyze the solution of the homogeneous set of equations. If we focus only at the solutions which do not decay in time, namely
%$|E_+|^2+|E_-|^2=const(t)$, it is easy to show that the condition $E_+=-E_-=E_0$ should be fulfilled. After substitution of this condition to the set of equation, the resulting nonlinear differential  equation can be solved yielding
\begin{align}
|E_\pm|=\sqrt{\frac{1}{3}(\delta n-\omega)}.
\end{align}
Therefore, in the case of Kerr-type focusing nonlinearity, BIC exists at all the frequencies smaller than the frequency of the linear BIC ($\omega<\delta n$). This is the manifestation of {\em the self-tuning of BIC}. If we pump the sufficient energy density in the system, namely $W=d^{-1}\int  |E(z)|^2dz=(1/3)(\delta n-\omega)$ then the resonant frequency is tuned such that the BIC coincides with pump frequency. This can be regarded as a way to excite BIC: we pump the system with non-resonant radiation, and the systems adjusts itself due to the nonlinearity to form a BIC. This type of behavior of nonlinear BIC has been previously discussed in~Refs.~\cite{Bulgakov2015,Pichugin2016} for nonlinear Fabry-Perot resonator and nonlinear impurity model.

We then look for the solutions of the inhomogeneous set of equations in the form $E_{\pm}=-ia_{\pm}e^{i\phi_0-i\delta\omega t}$. The resulting set of nonlinear algebraic equations has three classes of solutions. The first one is the symmetric solution $a_{+}=a_-=a_s$. This class of solutions generates a standard S-shaped bistability curve and will be treated numerically in Sec.~\ref{sec:num_sim}. The second class is asymmetric solutions, which have the form $a_{\pm}=a_0e^{i\phi_{\pm}}$, where $a_0$ does not depend on $\tilde I_p$. These solutions are written as:
\begin{align}
&a_{0}=\sqrt{\frac{1}{3}(\delta n-\delta \omega)},\\
&\tan\left(\frac{\phi_++\phi_-}{2}\right)=-\gamma/\delta n,\\
&\cos\left(\frac{\phi_+-\phi_-}{2}\right)=\frac{\sqrt{\gamma I_p}}{2a_0\sqrt{\gamma^2+\delta n^2}}.
\end{align}
It can be seen, that as $I_p$ approaches zero, the solution approaches the fully antisymmetric solution $\phi_+-\phi_-=\pi$. Moreover, as $I_p=\frac{4}{3\gamma}(\delta n-\delta\omega)(\gamma^2+\delta n^2)$, the solution becomes fully symmetric. A similar type of solutions, for which the pump power affect only the phase shift has been explored in ~\cite{Bulgakov2011a} for the case of nonlinear impurity model. It was termed~\textit{Josephson-like} current, since like in Josephson effect, the sine of the phase difference is proportional to the current (pump intensity in our case). 
Note, that these solutions may be stable in finite-size system as we will show in Section~\ref{sec:stability}.
Finally, there is a class of asymmetric solutions for which $|a_+|\neq |a_-|$, and $\phi_+=\pi+\phi_-$. These solutions will be treated numerically in the next section. 

\section{Numerical simulation \label{sec:num_sim}}
In this section, we carry out the numerical study of coupled-mode equations~\eqref{eq:CMT-1} and the reflection coefficient, which can be written as
\begin{equation}
	R = \left|r_{\text{slab}} + \sqrt{\frac{\tilde{\gamma}}{\tilde{I}_p}} \tilde{E}_{+} e^{i\phi_0} + \sqrt{\frac{\tilde{\gamma}}{\tilde{I}_p}} \tilde{E}_{-} e^{i\phi_0} \right|^2.
	\label{eq:R}
\end{equation}
\begin{figure}[htbp]
	\centering
	\includegraphics[width=\linewidth]{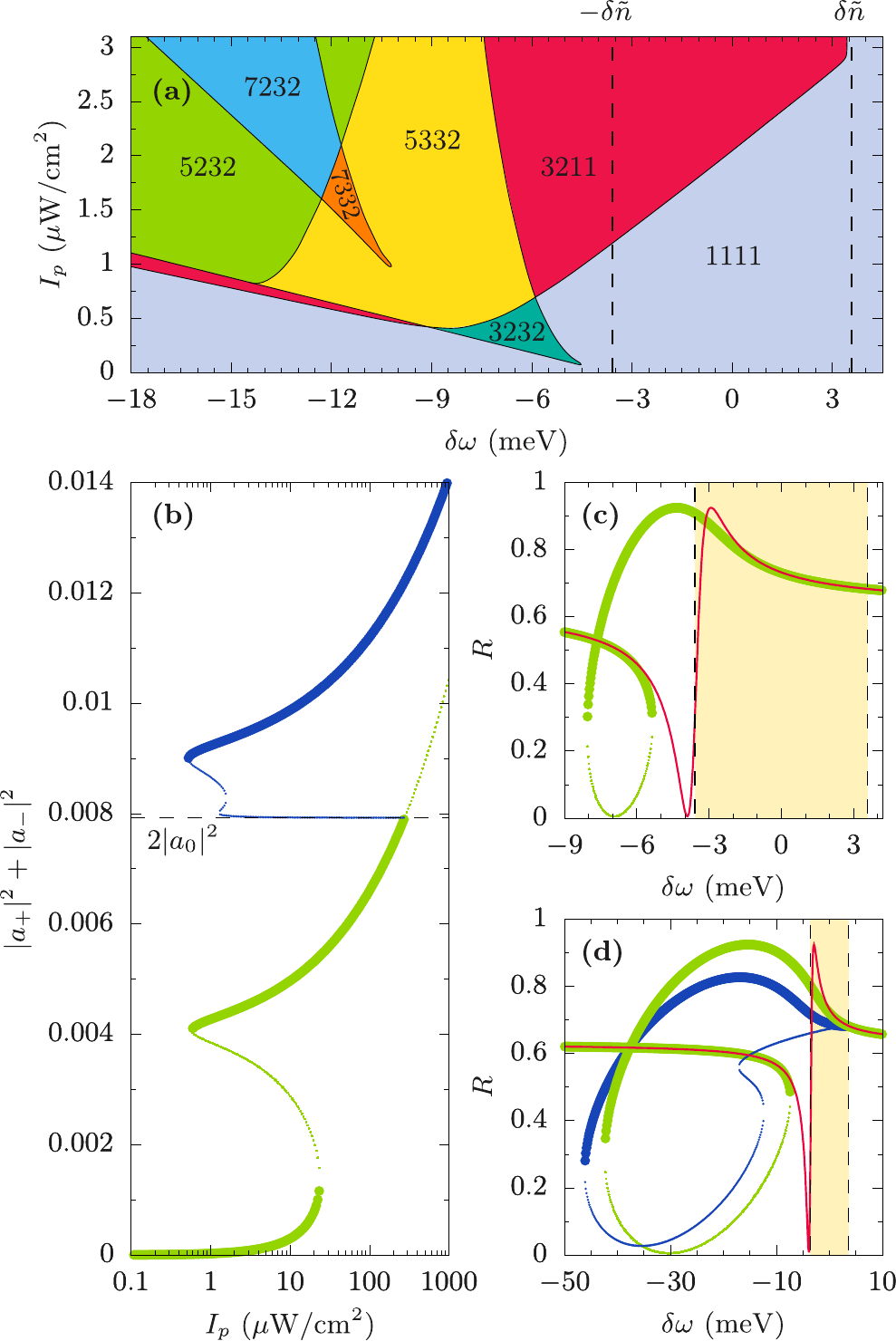}
	\caption{Graphical representation of the solutions of coupled-mode equations~\eqref{eq:CMT-1}. The four-digit numerical indices $absd$ on panel (a) have the following meaning: \textit{a} shows the total number of solutions, \textit{b} denotes how many of them are stable, \textit{c} is the number of symmetric solutions and \textit{d} is the number of stable symmetric solutions.
    The total field as the function of pump intensity is shown on (b) where $\delta\omega = -11.16$~meV. Note that the horizontal axis there has logarithmic scale. (c) and (d) shows the reflection coefficient of the structure with respect to detuning $\delta{\omega}$ at $I_p \sim 0.34~\mu$W/cm$^2$ on (c) and $I_p \sim 3~\mu$W/cm$^2$ on (d). All other parameters are the same for all panels and are given in the main text. In panels (c)-(d), the green dots correspond to symmetric solutions, the blue dots correspond to asymmetric and red lines indicate the linear solutions. The stable (unstable) solutions are shown by thick (tiny) dots. The dashed line in panel (b) corresponds to antisymmetric solutions $2|a_0|^2$. Shaded area on (c) and (d) indicates the region $-\delta\tilde{n} < \delta\tilde{\omega} < \delta\tilde{n}$. }
	\label{fig:solutions}
\end{figure}

For the numerical calculations we use the following set of parameters: we considered a silicon slab waveguide in vacuum  ($n_0 = 3.48$, $n_2 = 3 \cdot 10^{-18}~$m$^2$/W~\cite{lin2007dispersion, bristow2007two}). The depth of the etched grating was chosen to be $10$~nm while the thickness of the core layer without grating is $h = 100$~nm, the refractive index modulation amplitude in this case is $\delta n \approx 0.0316$. The lattice period $d$ was chosen in such a way that $\beta = 2\pi/d$ at wavelength $\lambda = 1~\mu$m. We have also added the material losses to the system by introducing additional diagonal losses term $\delta\gamma=0.25\gamma$ to matrix $\hat{\Gamma}$. The results of the numerical modeling are shown in Fig.~\ref{fig:solutions}.

Figure~\ref{fig:solutions}(a) shows the phase diagram depending on the pump detuning and intensity. The four numbers labeling different phases correspond to the number of stationary solutions, number of stable solutions, number of symmetric solutions, and number of stable symmetric solutions, respectively. Here, stability was checked with respect to homogeneous perturbations. It has been anticipated that nonlinear BIC can be destroyed by the modulation instability. It has been even suggested to exploit the instability to generate the frequency combs using BIC~\cite{Pichugin2015}. It can be seen that for the large positive detunings $\delta\omega>\delta n$, the systems supports a single symmetric stable solutions. In the region $-\delta n <\delta \omega <\delta n$ and large enough intensities, another phase exists where two additional asymmetric solutions emerge, one of which is stable. Therefore, the nonlinearity could result in {\em breaking of the symmetry protected BIC} and its transformation into symmetric solution with radiation losses. 

The situation complicates drastically as detuning becomes less than $-\delta n$. Namely, already at moderate pumping intensities, a multistable behavior is observed with two symmetric and one asymmetric stable solution. We plot the dependence of the modes intensity on the pumping power for $\delta\omega = -11.16$~meV 
in Fig.~\ref{fig:solutions}(b). The symmetric solutions are shown in green, and they exhibit conventional bistable behavior. At the same time, we observe the additional non-symmetry branch shown in blue. Remarkably, in the absence of material losses ($\delta\gamma=0$) the blue branch starts from the zero intensity manifesting the fully antisymmetric BIC. It then splits into the horizontal branch which connects it with the symmetric solution and the branch of asymmetric solutions, which exhibits S-shape behaviour. 

The asymmetric solutions generate the energy flow in the plane of the waveguide $S_z$, which is proportional to $|a_+|^2-|a_-|^2$. Since the asymmetric solutions appear in pairs equivalent up to interchange of $a_+$ and $a_-$, the specific current direction is defined by the initial conditions. The spectra of the reflection coefficient at different pump intensities are shown in Figs.~\ref{fig:solutions}(c,d) and are compared with the linear reflection coefficient (thin red line). 

%The asymmetric solutions generate the energy flow in the plane of the waveguide $S_z=\frac{1}{2}\mathrm{Re}\left[E(z,t)\times (-i/k_0)(\partial/\partial z E(z,t))^*\right]\sim |a_+|^2-|a_-|^2$. Since the asymmetric solutions appear in pairs equivalent up to interchange of $a_+$ and $a_-$, the specific current direction is defined by the initial conditions. The spectra of the reflection coefficient at different pump intensities are shown in Figs.~\ref{fig:solutions}(c,d) and are compared with the linear reflection coefficient (thin red line). 

\section{Modulation instability}\label{sec:stability}
\begin{figure}[htbp]
	\centering
	\includegraphics[width=\linewidth]{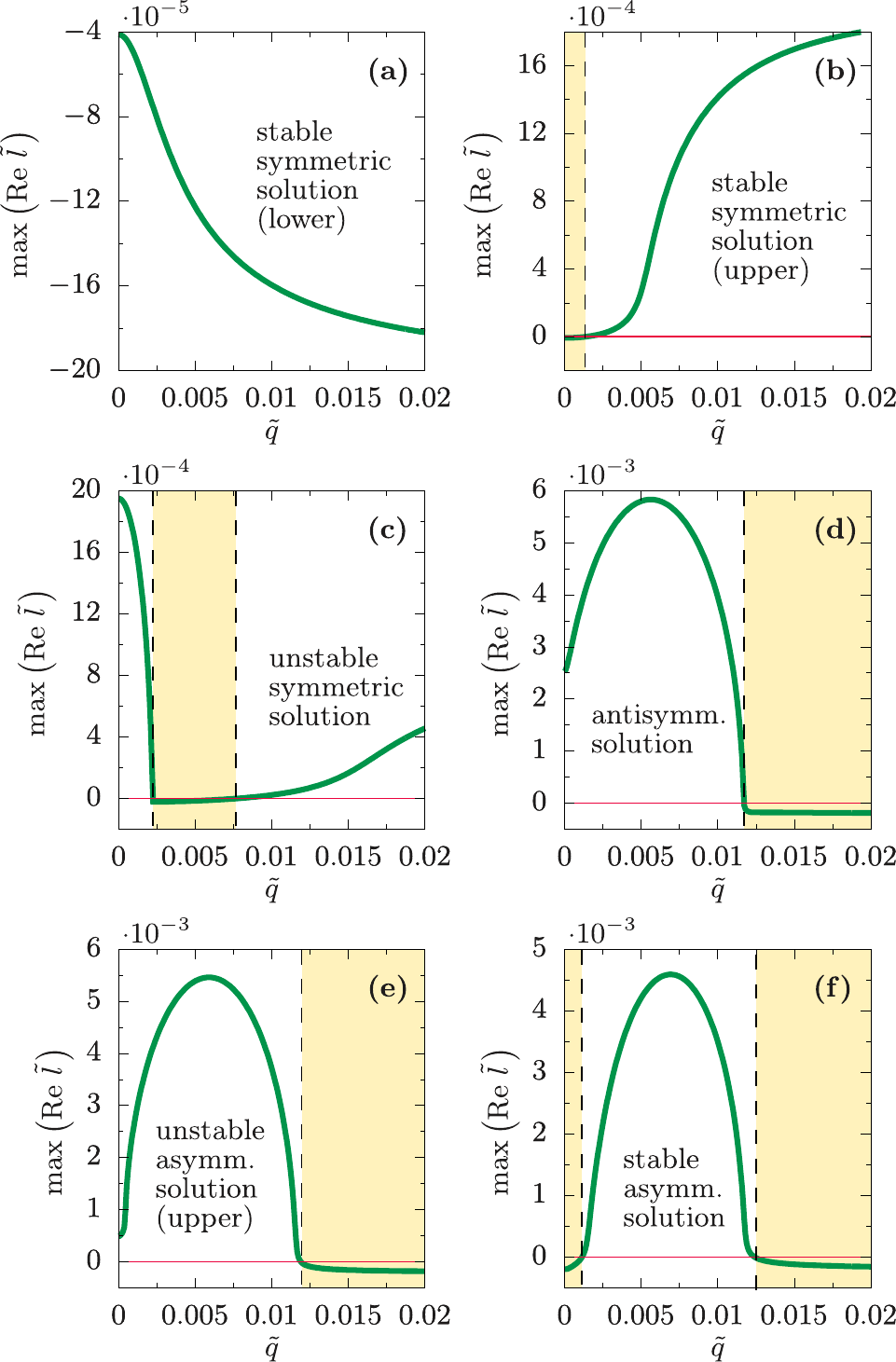}
	\caption{Stability analysis of solutions. Shaded area corresponds to stable solutions. For each plot $I_p \approx 1.5~\mu$W/m$^2$ and $\delta\omega = -11.16$~meV. All other parameters are given in the main text. }
	\label{fig:stability}
\end{figure}
As the next step, we analyze stability of the obtained solutions with respect to the longitudinal perturbations. For that we write the electric field in the form
\begin{equation}
	\tilde{E}_{\pm}(\tilde{z}, \tilde{t}) = \left(a_{\pm} + \epsilon f_{\pm}e^{i\tilde q\tilde z} e^{\tilde l \tilde{t}} + \epsilon g_{\pm}e^{-\tilde q \tilde z} e^{\tilde l^{*} \tilde t} \right) e^{-i \delta\tilde{\omega} \tilde{t}},
\end{equation}
where $\epsilon$ is the amplitude of the small perturbation, $\tilde q$ is the wavevector of the perturbation, which is a parameter, and $ \tilde l$ is the complex eigenfrequency. We substitute the expression for the electric field to the initial differential equations and linearize them with respect to $\epsilon$. The solution of the resulting eigenvalue problem for $\tilde l$, gives the spectrum of the linear perturbations. If at least one of the eigenvalues has positive real part, then our initial solution is unstable. Figure~\ref{fig:stability} shows the spectra of the linear perturbations with largest real part of $\tilde l$ for different classes of stationary solutions. The stability with respect to homogeneous perturbations labeled in Fig.~\ref{fig:solutions} corresponds to the case of $\tilde q=0$. We can see that while the lower stable branch of the symmetric solution is stable for all possible $\tilde q$ [Fig.~\ref{fig:stability}(a)], the upper branch is unstable to the perturbations with any finite $\tilde q$ larger than some critical $\tilde q_{crit}$ [Fig.~\ref{fig:stability}(b)]. It has been shown previously for the similar system that the instability of the upper symmetric branch leads to the soliton formation~\cite{Yulin2005}. At the same time, we can see that all of the asymmetric solutions, are stable with respect to perturbations with wavevectors, larger than some finite critical value $\tilde q_{max}$ as can be seen in Figs.~\ref{fig:stability}(d-f). This has interesting consequences for the stability conditions in finite structures. Namely, since only perturbations which have wavelength smaller than the system size may exist, the perturbations with $\tilde q<1/N$ (where $N$ is the number of periods) will decay. Thus, if the system size is less than $\tilde q_{max}$, it will be linearly stable with respect to longitudinal perturbation.   Therefore, a finite size of the structure could stabilize the solutions, which are unstable in the infinite system.
%It should be noted, thought that true BIC exists only in the infinite structure, and only a so-called quasi-BIC state with finite Q-factor may exist in finite structures.

\section{Conclusion}

To conclude, we have demonstrated the existence of the nonlinear BIC in a simple structure,  periodically corrugated silicon waveguide. We have shown, how the existence of the BIC leads to the emergence of the multistable behavior in the structure at moderate pump intensities. Moreover, it has been shown that finite system size may stabilize the solutions, which are unstable in the infinite system. The moderate level of pump intensities required for the optical switching in the structures supporting BIC states opens new avenues for the realization of all-optical switchers exploiting bound states in the continuum.
\begin{acknowledgements}
This work is supported by RSF (17-12-01581).
\end{acknowledgements}

% Bibliography
\bibliography{manuscript}

\end{document}